\documentclass{isprs} 
\usepackage{subfigure}
\usepackage{setspace}

\usepackage{geometry} 
\usepackage{epstopdf}
\usepackage[labelsep=period]{caption}  
\usepackage[british]{babel} 
\usepackage[hang]{footmisc}
\usepackage[hidelinks]{hyperref} 
\hypersetup{colorlinks=false}
\usepackage{marvosym} 
\usepackage{enumitem} 
\usepackage{tabularray} 
\setlist[itemize]{noitemsep, topsep=0pt} 
\setlist[enumerate]{noitemsep, topsep=0pt} 
\setlist[description]{noitemsep, topsep=0pt} 
\usepackage[T1]{fontenc} 
\usepackage{float} 
\usepackage[labelfont=bf]{caption} 

\begin{document}
\title{Impact of geolocation data on augmented reality usability: \\a comparative user test}
\date{}

\author{
 J. Mercier\textsuperscript{1,2}\thanks{Corresponding author} , N. Chabloz\textsuperscript{1}, G. Dozot\textsuperscript{1}, C. Audrin\textsuperscript{3}, O. Ertz\textsuperscript{1}, E. Bocher\textsuperscript{2}, D. Rappo\textsuperscript{1}
}

\address{
    \textsuperscript{1 }
    Media Engineering Institute (MEI), School of Engineering and Management Vaud, HES-SO, Yverdon-les-Bains, Switzerland - \\(julien.mercier, nicolas.chabloz, gregory.dozot, olivier.ertz, daniel.rappo)@heig-vd.ch\\
	\textsuperscript{2 }Lab-STICC, UMR 6285, CNRS, Université Bretagne Sud, Vannes, France - julien.mercier@univ-ubs.fr, erwan.bocher@cnrs.fr\\
     \textsuperscript{3 }University of Teacher Education, HES-SO, Lausanne, Switzerland - catherine.audrin@hepl.ch\\
}


\commission{XX, }{YY} 
\workinggroup{XX/YY} 
\icwg{}   

\abstract{
While the use of location-based augmented reality (AR) for education has demonstrated benefits on participants’ motivation, engagement, and on their physical activity, geolocation data inaccuracy causes augmented objects to jitter or drift, which is a factor in downgrading user experience. We developed a free and open source web AR application and conducted a comparative user test (n~=~54) in order to assess the impact of geolocation data on usability, exploration, and focus. A control group explored biodiversity in nature using the system in combination with embedded GNSS data, and an experimental group used an external module for RTK data. During the test, eye tracking data, geolocated traces, and in-app user-triggered events were recorded. Participants answered usability questionnaires (SUS, UEQ, HARUS). We found that the geolocation data the RTK group was exposed to was less accurate in average than that of the control group. The RTK group reported lower usability scores on all scales, of which 5 out of 9 were significant, indicating that inaccurate data negatively predicts usability. The GNSS group walked more than the RTK group, indicating a partial effect on exploration. We found no significant effect on interaction time with the screen, indicating no specific relation between data accuracy and focus. While RTK data did not allow us to better the usability of location-based AR interfaces, results allow us to assess our system’s overall usability as excellent, and to define optimal operating conditions for future use with pupils. }

\maketitle


\section{Introduction}\label{introduction}
\sloppy
This study is part of the ongoing \textit{BiodivAR} project, which attempts to assess the potential benefits of using augmented reality (AR) for outdoor education on biodiversity. In AR interfaces, digital objects can be overlaid on top of users’ field of view in real-time, through the screen of a mobile device or a head-mounted display. When used sensibly in an educational setting, it may convey the impression of an enriched environment and make the material more attractive, thus motivating students to learn~\cite{geroimenko_augmented_2020,alnagrat_review_2022}. The most reported positive effects of AR in education are learning gains and motivation~\cite{bacca_augmented_2014}. Our research is focused on the use of \textit{location-based} AR in particular, where the position of augmented objects is computed based on their geographic coordinates relative to the user’s location as estimated by the mobile device’s GNSS. With this technology, augmented objects can be built remotely from any given geodata, as opposed to marker-based AR which requires physical markers to be physically placed on target locations. Location-based AR specially promotes learning in context~\cite{arvola_mobile_2021, chiang_augmented_2014}, ecological engagement~\cite{bloom_promoting_2010}, and causes users to experience a positive interdependence with nature~\cite{oshea_research_2011}, which fosters improved immersion and learning. Last but not least, location-based AR shows positive effects on the physical activity of users across genders, ages, weight status, and prior activity levels~\cite{rauschnabel_adoption_2017}. However, location-based AR requires steady and continuously accurate data to operate. While GNSS technology has evolved and improved in the past decades, it has been more of an evolution than a revolution. Usability issues have been reported by a number of studies~\cite{chiang_augmented_2014,dunleavy_affordances_2009,ryokai_off_2013,admiraal_concept_2011,lee_cityviewar_2012}, most of which blame the inaccuracy of mobile devices’ embedded GNSS sensors. Some studies considered that these recurring problems made AR distracting and frustrating and eventually favored marker-based AR, which is more advanced and offers better user experience~\cite{bressler_mixed_2013,debandi_enhancing_2018}. 

\section{Background}\label{background}
A first proof-of-concept was developed in 2017, featuring a series of geolocated points of interest (POIs) on biodiversity. A test with ten-year-old pupils confirmed the relevance of using AR to support educational field trips~\cite{ingensand_augmented_2018} while also revealed usability challenges: 
\begin{enumerate}[noitemsep,topsep=0pt,leftmargin=1em]
\setlength\itemsep{0em}\setlength\parskip{0em}\setlength\topsep{0em}\setlength\partopsep{0em}\setlength\parsep{0em} 
    \item{The system should allow non-expert users to create AR experiences~\cite{cubillo_preparing_2015}} 
    \item{Users should be able to publish observations rather than being restricted to a passive viewing role;}
    \item{The instability of augmented objects deteriorates usability. Participants spent 88.5~\% of the time looking at the tablet rather than with the surrounding nature. This imbalance could be in part related to inaccurate geolocation data: participants were observed spending considerable time reorienting themselves~\cite{ingensand_augmented_2018}.} 
\end{enumerate}

\begin{figure}[ht!]
    \centering
    \includegraphics[width=1.0\columnwidth]{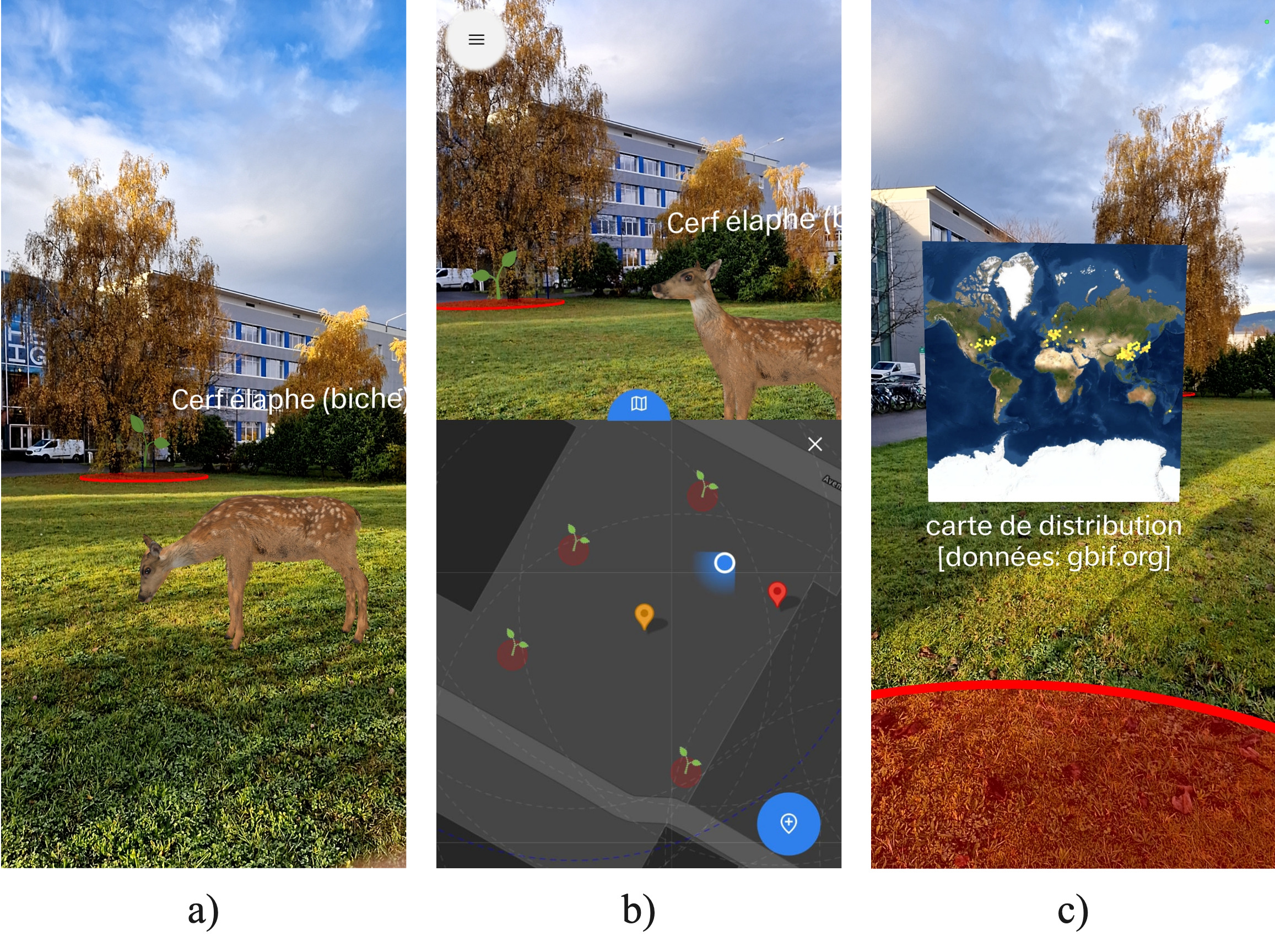}
    	\caption{\textit{BiodivAR}’s AR interface: a) view of two POIs from a distance; b) the 2D map is opened in split view; c) after entering the radius of a POI, contextual data on the adjacent plant specimen is triggered. \url{https://biodivar.heig-vd.ch/}}
    \label{fig:interface}
\end{figure}

\begin{figure}[ht!]
    \centering
    		\includegraphics[width=1.0\columnwidth]{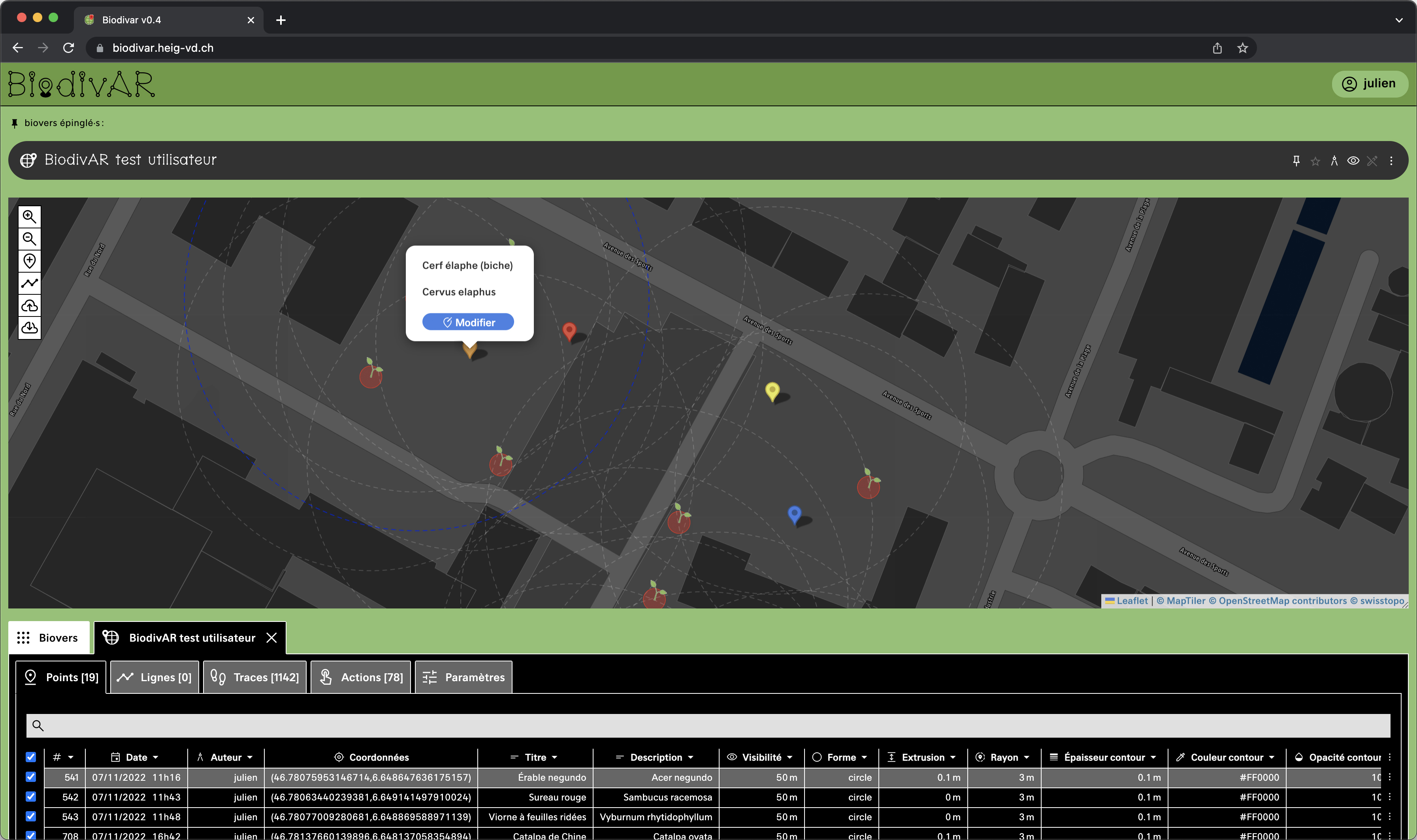}
    	\caption{\textit{BiodivAR}’s cartographic authoring tool.}
    \label{fig:authoring-tool}
\end{figure}

In order to address these identified issues, we developed \textit{BiodivAR}\footnote{The web application is released under the GNU General Public License v3.0. It is accessible (no download required) at: \url{https://biodivar.heig-vd.ch}. The source code is available at \url{https://github.com/MediaComem/biodivar}.}, a free and open source (GNU GPLv3.0) web application using a user-centered design process~\cite{mercier_biodivar_2023}. It was built using the web framework A-Frame\footnote{\url{https://github.com/aframevr/aframe} (MIT License)}, for which we also created a custom library\footnote{https://github.com/MediaComem/LBAR.js/ (MIT License)} for the creation of WebXR location-based objects in A-Frame. We used the Leaflet\footnote{\url{https://github.com/Leaflet/Leaflet} (FreeBSD License)} library for the interactive maps. \textit{BiodivAR} enables the creation and visualization of geolocated POIs in AR (see Figure~\ref{fig:interface}) as well as a cartographic authoring tool for the collaborative management of AR environments (see Figure~\ref{fig:authoring-tool}). They can be shared publicly with or without editing privileges. The application allows anyone without technological know-how to create AR environments by importing/exporting geospatial data and styling POIs by attaching medias to them. Medias can be location-triggered (visible/audible) according to various distance thresholds set by the author.  

\section{Research goals}\label{research-goals}

The purpose of our research \textit{overall} is to assess the potential benefits of using this application in the context of biodiversity education. Before introducing the tool to pupils, it seemed important to ensure its usability. This comparative user test will allow us to define and guarantee the best possible conditions of use for a younger audience. The goals of this study can be synthesized as follows: 
\begin{enumerate}[noitemsep,topsep=0pt,leftmargin=1em]
\setlength\itemsep{0em}\setlength\parskip{0em}\setlength\topsep{0em}\setlength\partopsep{0em}\setlength\parsep{0em} 
    \item{Assess the overall usability of the AR application.} 
    \item{Assess the impact of geolocation data accuracy on usability, exploration, and focus.}
    \item{Gather user feedback for future improvements\footnote{The qualitative feedbacks were not included in this paper, as we exclusively focused on the quantitative data and group comparison.}.}
\end{enumerate}

The literature review and the observations made during the first iteration led us to propose the following hypothesis: Inaccurate geolocation data negatively affects usability. Additionally, we are looking to investigate the impact that geolocation data accuracy may have on exploration and focus in location-based AR, about which we have not been able to find any literature. The resulting research questions are:

\begin{itemize}
 \setlength\itemindent{-2em}\setlength\itemsep{0em}\setlength\parskip{0em}\setlength\topsep{0em}\setlength\partopsep{0em}\setlength\parsep{0em} 
    \item[]{Q1: Does geolocation data accuracy predict usability scores?}
    \item[]{Q2: Is geolocation data accuracy related to exploration?\footnote{Exploration is represented by the distance walked, the number of POIs visited, and the number of times the 2D map was opened.}} 
    \item[]{Q3: Is geolocation data accuracy related to focus?\footnote{Focus is represented by the ratio of time spent gazing at the screen \textit{versus} the real world.}} 
\end{itemize}

\section{Materials and methods}\label{materials-and-methods}

    \subsection{Experimental design}\label{sec:experimental-design}
\begin{figure}[ht!]
        \centering
        		\includegraphics[width=1.0\columnwidth]{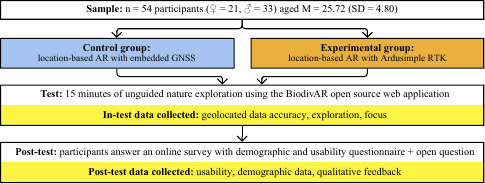}
        	\caption{Experimental design of the comparative user test.}
        \label{fig:experiment-design}
    \end{figure}

    The present study aims to measure and compare the usability of a location-based AR application used in combination with different geolocation data sources. Using our authoring tool, we created an AR environment with POIs on biodiversity in the surroundings of the School of Engineering and Management Vaud in Yverdon-les-Bains (Switzerland). After a brief introduction to the tool, all participants freely explored the AR environment for 15 minutes using a Samsung Galaxy Tab Active3 tablet with a SIM card for cellular data. As shown in Figure~\ref{fig:experiment-design}, the comparative user test (n = 54) includes in two groups: 
    \begin{description}
    \setlength\itemindent{-1.5em}\setlength\itemsep{0em}\setlength\parskip{0em}\setlength\topsep{0em}\setlength\partopsep{0em}\setlength\parsep{0em} 
        \item[GNSS]{the control group received geolocation data coming from the GNSS sensor embedded in the mobile device} 
        \item[RTK]{the experimental group received geolocation data coming from an external Ardusimple RTK kit\footnote{\url{https://www.ardusimple.com/product/rtk-handheld-surveyor-kit/}}.}
    \end{description}
    
    \subsection{Participants}\label{sec:participants}
    The sample includes 54 participants (\Female \;= 21, \Male \;= 33), with a mean age of M~=~25.72 (SD~=~4.80). They are students and collaborators of the School of Engineering and Management Vaud, and they each signed an informed consent form for the use of the data collected. Login credentials (identifier + password) were created for each participant to record their data separately and facilitate comparison. Among them, 47 agreed to wear eye-tracking glasses, of which 41 successfully recorded data. They were randomly assigned to each group. The control group’s (GNSS) mean age is M~=~27.5 (SD~=~6.09), and it includes 12 \Female \;and 15 \Male. The experimental group’s (RTK) mean age is M~=~24.2 (SD~=~2.22) and it includes 9 \Female \;and 18 \Male. The first participant eventually had to be excluded from the final results because they experienced numerous crashes due to a bug that was fixed for the subsequent participants. The treatment they received was therefore too different to compare. 
    
    \subsection{Data collection and processing}\label{sec:data-collection}
    The four main concepts our study seeks to connect are “location data accuracy”, “usability”, “exploration”, and “focus”. The measurable observations we chose to represent those concepts are listed in Table~\ref{tab:operationalization-table}. In our experiment, the two groups (or treatments) operationalize the concept of “geolocation data accuracy”. This concept is represented by two variables: \textit{accuracy} and \textit{continuity}. The accuracy attribute is provided by the Geolocation API along with the horizontal location data as latitude and longitude\footnote{\url{https://w3c.github.io/geolocation-api}}. It denotes the accuracy level of the latitude and longitude coordinates in meters. We use the average accuracy participants were exposed to while in AR mode as the indicator for accuracy. However, in the specific context of location-based AR, sudden changes in data accuracy heavily impact the display of augmented objects in the interface. An indicator for continuity in the data is thus the amount of outliers–i.e. the points that are visibly out of a user’s trajectory (as shown in Figure~\ref{fig:geolocated-traces-and-events}). An additional indicator for continuity in the data is the standard deviation of the data accuracy the participants of each group was exposed to. As far as the concept of “usability” goes, it is represented by a series of nine variables whose indicators are the different scales of the three questionnaires (SUS, HARUS, UEQ):  \textit{overall usability}, \textit{ease of handling}, \textit{ease of understanding}, \textit{attractability}, \textit{user-friendliness}, \textit{efficiency}, \textit{dependability}, \textit{motivation}, \textit{innovativeness}. The concept of “exploration” is represented by three variables: \textit{quantity}, \textit{diversity}, and \textit{ease}. The distance walked is the indicator of the quantity of exploration. The amount of POIs visited is the indicator of the diversity of exploration. An important use of the 2D map may indicate that participants required assistance in navigating. The amount of times the 2D map was opened is thus the indicator of the ease users had exploring. Finally, the concept of “focus” in our study is represented by a \textit{screen interaction} variable, whose indicator is the amount of time participants spent interacting with the tablet screen \textit{versus} with the real world.  
    
    \begin{table}[h]
    \centering
    \renewcommand{\baselinestretch}{0.9} 
        \begin{tblr}{rows = {abovesep=0pt,belowsep=0pt},row{1-30}={font=\scriptsize},colsep=5pt,hlines,vlines}
            \textbf{Concept}&\textbf{Variable}&\textbf{Indicator}\\
            \SetCell[r=3]{}{Geolocation\\data accuracy} & Quality & Average geolocation data accuracy\\
             &\SetCell[r=2]{}{Continuity} & Amount of outliers\\
             & & Standard deviation of data accuracy \\
            \SetCell[r=9]{}Usability & Overall usability & SUS score\\
             & Ease of handling & HARUS (manipulability) score\\
             & Ease of understanding & HARUS (comprehensibility) score\\
             & Attractability & UEQ (attractiveness) score\\
             & User-friendliness & UEQ (perspicuity) score\\
             & Efficiency & UEQ (efficiency) score\\
             & Dependability & UEQ (dependability) score\\
             & Motivation & UEQ (stimulation) score\\
             & Innovativeness & UEQ (novelty) score\\
            \SetCell[r=3]{}Exploration & Quantity & Distance walked\\
             & Diversity & Amount of POIs visited\\
             & Ease & Amount of times 2D map was opened\\
            Focus & Screen interaction & Interaction time with tablet screen
        \end{tblr}
    \caption{Operationalization table.}
    \label{tab:operationalization-table}
    \end{table}
    
        \subsubsection{Geolocation data accuracy}\label{sec:geolocation-data}
\begin{figure}[ht!]
            \centering
            \includegraphics[width=1.0\columnwidth]{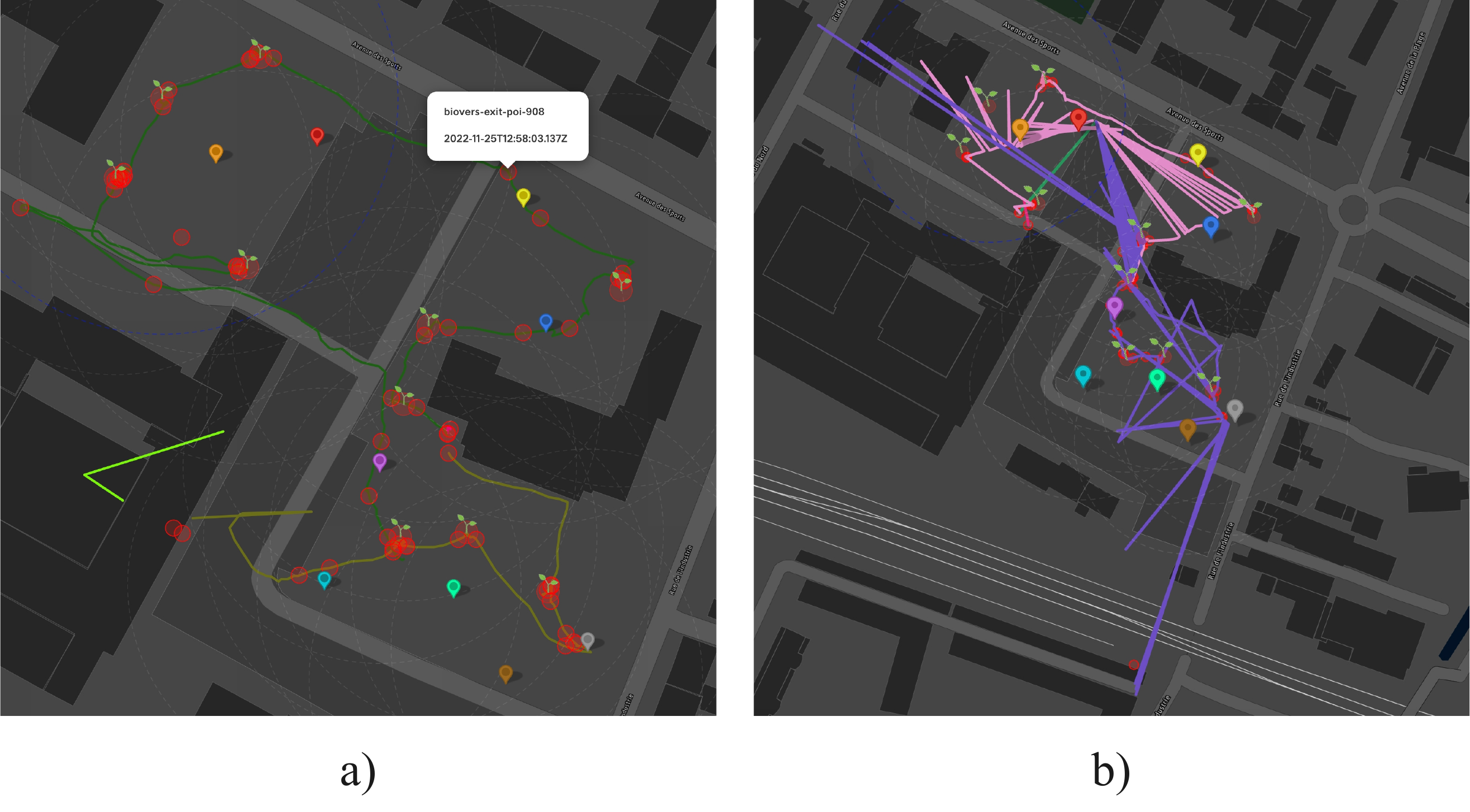}
            \caption{a) A trajectory from the GNSS group. The short light green line is at an impossible location (on top of a tall building), indicating outliers. b) A trajectory from the RTK group. The star-shaped spikes indicate the presence of many outliers.} 
            \label{fig:geolocated-traces-and-events}
        \end{figure}

        During the test, participants’ geographical coordinates were logged at 1 Hz. Each log also contains an attribute for location accuracy, user ID and a timestamp. The resulting users’ trajectories can be visualized in the application (see Figure~\ref{fig:geolocated-traces-and-events}) and downloaded as GeoJSON files for further analysis. The color of the trajectory changes when the AR session is stopped and resumed again. We downloaded the data and calculated the mean location accuracy each participant was exposed to. As shown in Figure~\ref{fig:geolocated-traces-and-events}, the trajectories–in particular that of the RTK group–contained outliers, which were removed manually using the free and open source software QGIS to get a more accurate estimate of the actual distance travelled (as an indicator of our “exploration quantity” variable, see~\ref{sec:exploration}). By calculating the different amount of points before and after this manual processing, the outliers were summed for each participant. Once the data was cleaned, we calculated the total distance walked by each participant. Because there were variations in the duration of each participant’s test (min = 9$'$14, max = 24$'$11 s), the data was normalized for a duration of 15 minutes. This allowed us to calculate: 
        \begin{enumerate}[noitemsep,topsep=0pt,leftmargin=1em]
        \setlength\itemsep{0em}\setlength\parskip{0em}\setlength\topsep{0em}\setlength\partopsep{0em}\setlength\parsep{0em} 
            \item The average geolocation data accuracy
            \item The amount of outliers in the data
            \item The standard deviation of the geolocation data accuracy
        \end{enumerate}
            
        \subsubsection{Usability}\label{sec:usability}
        Immediatly after the test, participants answered an online survey containing demographic questions (age, gender), an open question for qualitative feedback, and three usability questionnaires: 
        \begin{itemize}[noitemsep,topsep=0pt,leftmargin=1em]
        \setlength\itemindent{0em}\setlength\itemsep{0em}\setlength\parskip{0em}\setlength\topsep{0em}\setlength\partopsep{0em}\setlength\parsep{0em} 
            \item SUS (System Usability Scale) is a generic, technology-independent 10 item questionnaire on a 5 point Likert scale, frequently used for generic evaluation of a system~\cite{brooke_sus_1996}. The Cronbach’s alpha of the SUS questionnaire is 0.79, showing an appropriate internal consistency. In accordance with the instructions of the scale’s authors, the SUS score is calculated as follows: 1 point was subtracted from the odd-numbered (phrased positively) items’ scores. We subtracted the even-numbered (phrased negatively) items score to 5. The processed scores were added together and then multiplied by 2.5 to get an individual user’s score on a scale of 100. While a comparison between two scores is self-explanatory, we used an adjective scale ~\cite{bangor_determining_2009} to qualify the results individually. 
    
            \item HARUS (Handheld Augmented Reality Usability Scale) is a mobile AR-specific 16 item questionnaire~\cite{santos_usability_2014} on a 7 point Likert scale that focuses on handheld devices and emphasizes perceptual and ergonomic issues. The Cronbach’s alpha of the HARUS questionnaire is 0.798, showing appropriate internal consistency. It has two components: \textit{manipulability}—the ease of handling the AR system, and \textit{comprehensibility}—the ease to read the information presented on screen. In accordance with the instructions of the scale’s authors, the HARUS scores are calculated as follows: We subtracted the odd-numbered (phrased negatively) items score to 7. 1 point was subtracted from the even-numbered (phrased positively) items’ scores. The processed scores for items 1 to 8 were added together, divided by 48, and multiplied by 100 to get the individual “manipulability” score on a scale of 100. Similarly, the processed scores for items 9 to 16 were added together, divided by 48, and multiplied by 100 to get the individual “comprehensibility” score on a scale of 100. HARUS was designed so that its scores are commensurable with SUS scores. 
            
            \item UEQ (User Experience Questionnaire) is a 26 item questionnaire 
            in the form of semantic differentials: each item is scored on a 7 point scale (from -3 to +3, with 0 as neutral) with two terms with opposite meanings at each extreme (i.e. attractive|unattractive). It provides a comprehensive measure of user experience~\cite{laugwitz_construction_2008}. It includes six scales, covering classical usability aspects such as \textit{efficiency} (can users solve their tasks without unnecessary effort?), \textit{perspicuity} (is it easy to learn how to use the application?), and \textit{dependability} (does the user feel in control of the interaction?), as well as broader user experience aspects such as \textit{attractiveness} (do users like the application?), \textit{novelty} (is the application innovative and creative?), and \textit{stimulation} (is it exciting and motivating to use the application?). UEQ is typically routinely used to statistically compare two version of a system to check which one has the better user experience. Thus, the UEQ evaluations of both systems or both versions of a system are compared on the basis of the scale means for Each UEQ scale. 
            \textit{Attractiveness} is calculated by averaging the scores from items 1, 12, 14, 16, 24, and 25. \textit{Perspicuity} is calculated by averaging the scores from items 2, 4, 13, and 21. \textit{Efficiency} is calculated by averaging the scores from items 9, 20, 22, and 23. \textit{Dependability} is calculated by averaging the scores from items 8, 11, 17, and 19. \textit{Stimulation} is calculated by averaging the scores from items 5, 6, 7, and 18. Novelty is calculated by averaging the scores from items 3, 10, 15, and 26. Values range between -3 (horribly bad) and +3 (extremely good), but in general only values in a restricted range will be observed. The calculation of means over a panel of participants make it extremely unlikely to observe values above +2 or below -2, as specified in the UEQ handbook~\cite{schrepp_user_2015}. As per their interpretation, values between -0.8 and 0.8 correspond to a neutral evaluation of the corresponding scale and values greater than 0,8 represent a positive evaluation. 
        \end{itemize}
    
        These questionnaires provided scores for the nine scales reported in Table~\ref{tab:operationalization-table} as indicators of our usability variables.

        \subsubsection{Exploration}\label{sec:exploration}
        During the test, various in-app, user-triggered events were recorded by the application. These included: when the AR session was initiated or exited, when the 2D map was opened or closed, and when the triggering radius of a POI was entered or exited. Each log also contains the coordinates the action took place at, the user ID and a timestamp. The resulting users’ action log can be visualized in the application and downloaded as GeoJSON files. Events are represented with red circles on the 2D map (see Figure~\ref{fig:geolocated-traces-and-events}). We downloaded the data and calculated the number of POIs each participant visited as well as how many times they opened the 2D map. These values (POIs visited, 2D map opened) were normalized for a test duration of 15 minutes. This allowed us to calculate: 
        \begin{enumerate}[noitemsep,topsep=0pt,leftmargin=1em]
        \setlength\itemsep{0em}\setlength\parskip{0em}\setlength\topsep{0em}\setlength\partopsep{0em}\setlength\parsep{0em}
            \item The amount of POIs visited
            \item The amount of times the 2D map was opened
        \end{enumerate}
        The distance walked by each participant was calculated from the geolocation data (see~\ref{sec:geolocation-data}).

        \subsubsection{Focus}\label{sec:focus}
\begin{figure}[H]
            \centering
            \includegraphics[width=1.0\columnwidth]{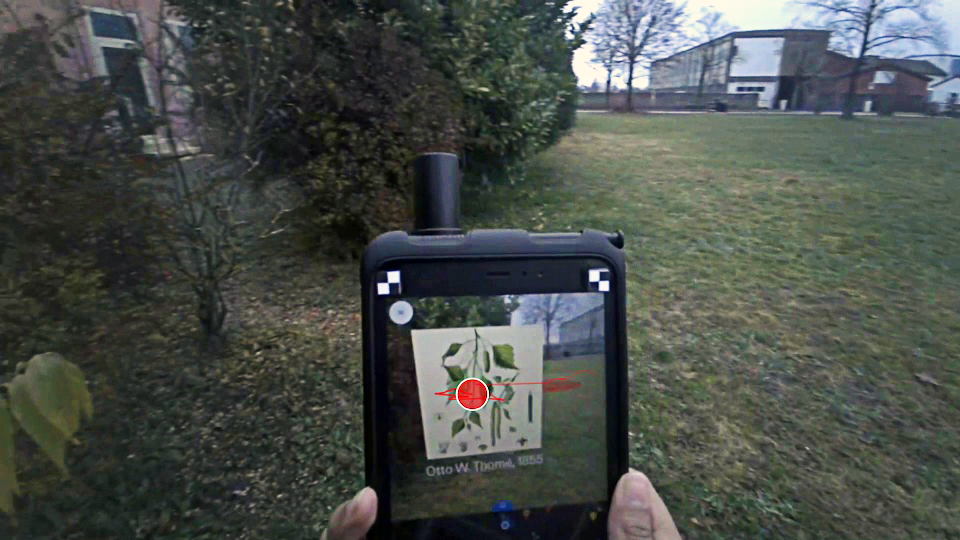}
            \caption{Eye tracking data sample. The user’s gaze is located within the tablet screen area.} 
            \label{fig:eye-tracking}
        \end{figure}

        The goal of using eye tracking glasses and data in our study is to determine for how long participants were looking in or out of the tablet screen. 47 out of 54 participants were able–and agreed–to wear eye trackers (Tobii Pro Glasses 3), recording their gaze for the duration of the test. The 7 participants that didn’t either choose not to or couldn’t because they had prescription glasses. Despite rigorous implementation, 6 recordings did not work as expected and no files were saved. The 41 remaining recordings were imported in Tobii’s analysis software. Unfortunately, its tools do not support tracking of moving areas of interest (i.e. the surface of the tablet). We exported the videos with the overlaying gaze point and extracted 10 frames per second, resulting in a dataset of 380K images, an instance of which is shown in Figure~\ref{fig:eye-tracking}. We attempted to classify the data with openCV pattern recognition, but the variability prevented from obtaining any results. We resolved to train a deep learning multiclass image classifier model by fine-tuning a pretrained vision transformer (ViT) model with our dataset~\cite{dosovitskiy_image_2020}. We first had to manually label a random selection of 10K frames with “in” or “out” labels corresponding to whether the point was in or out of the tablet screen (see Figure~\ref{fig:eye-tracking}). After training for only one epoch using Google’s colaboratory and obtaining a satisfying validity of ~95\%, we inferred the whole dataset which provided a label for every frame\footnote{The code used to fine-tune the ViT model is accessible in the following Jupyter Notebook: \url{https://colab.research.google.com/drive/1sYxbJQ-7FrScr7R87qwmqKZcb837LWhJ\#scrollTo=JLseEgvycDGy}. The dataset and the trained model are available here: \url{https://huggingface.co/julienmercier}.}. They were encoded in order to calculate the ratio of time each user spent looking at the tablet screen \textit{versus} outside of it, at the real world. 
    
\section{Results}\label{sec:results}

    \subsection{Data analysis}\label{sec:data-analysis}
    Statistical analysis were made with the free and open platform Jamovi~\cite{the_jamovi_project_jamovi_2022}. In the following subsections, we report descriptive statistics (M, SD), and compare our groups (GNSS \textit{versus} RTK) using an independant Student \textit{t}-test to emphasize to which extent both groups differ on our variables of interest. In cases where the homogeneity of variances assumption is not met, we used a Welch \textit{t}-test, which is more robust\footnote{The data is available here: \url{https://zenodo.org/record/7845707}.}. 
    
    \subsection{Geolocation data accuracy}\label{sec:geolocation-data-accuracy-results}
    
        \subsubsection{Average geolocation data accuracy}\label{sec:average-geolocation-data-accuracy} 
\begin{figure}[ht!]
            \centering
            \includegraphics[width=1.0\columnwidth]{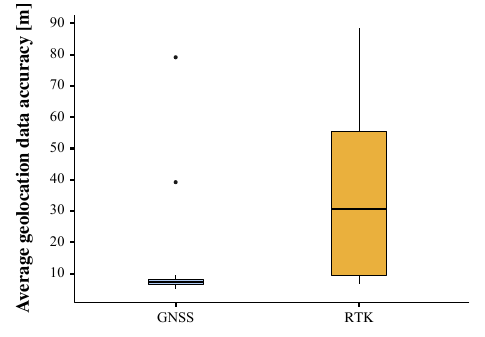}
            \caption{Geolocation data accuracy by group.}
            \label{fig:geolocation-data-accuracy-comparison}
        \end{figure}

As shown in Figure~\ref{fig:geolocation-data-accuracy-comparison}, the mean accuracy for the GNSS group is M~=~11.0 (SD~=~15.3), and M~=~33.6 (SD~=~24.8) for the RTK group. The value is in meters, meaning the data the GNSS group was exposed to was accurate within a 11 meters radius, whereas the RTK group got data accurate within a 33.6 meters radius. A Welch \textit{t}-test was used. The results show a significant difference between the two groups (t(43.5)~=~-3.99, p~=~<.001). 
    
        \subsubsection{Outliers}\label{sec:results-outliers} 
\begin{figure}[H]
            \centering
            \includegraphics[width=1.0\columnwidth]{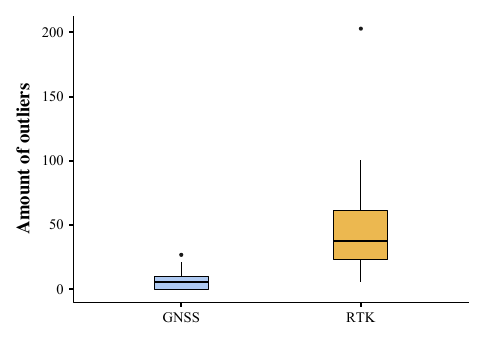}
            \caption{Amount of outliers by group.}
            \label{fig:outliers-comparison}
        \end{figure}

As shown in Figure~\ref{fig:outliers-comparison}, the GNSS group trajectories contained M~=~7.2 (SD~=~7.55) outliers, and these of the RTK group M~=~46.8 (SD~=~40.1). A Welch \textit{t}-test was used. The results show a significant difference between the two groups (t(27.9) = -5.04, p~=~<.001). 
        
        \subsubsection{Standard deviation geolocation data accuracy}\label{sec:results-sd-geolocation-data-accuracy} 
\begin{figure}[ht!]
            \centering
            \includegraphics[width=1.0\columnwidth]{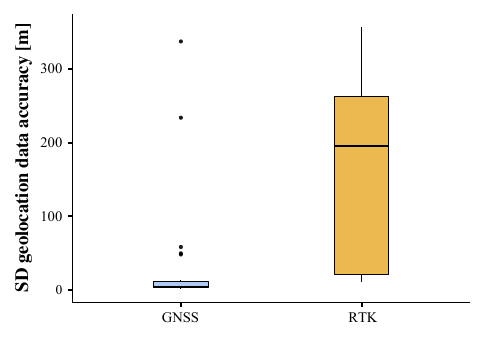}
            \caption{Standard deviation geolocation data accuracy by group.}
            \label{fig:SD-location-data-accuracy-comparison}
        \end{figure}

As shown in Figure~\ref{fig:SD-location-data-accuracy-comparison}, the data participants from the GNSS group were exposed to had a standard deviation of M~=~32.0 (SD~=~77.7), and that of the RTK group M~=~168.3 (SD~=~120.1). A Welch \textit{t}-test was used. The results show a significant difference between the two groups (t(44.7) = -4.93, p~=~<.001). 
            
    \subsection{Usability}\label{sec:usability}
    The means of each group for all nine scales from the three usability questionnaires are reported in Table~\ref{tab:usability-table} along with \textit{t}-test’s p values for significance assessment. 
    
    \begin{table}[h]
        \centering
        \renewcommand{\baselinestretch}{0.9} 
        \begin{tblr}{rows = {abovesep=0pt,belowsep=0pt},row{1-10}={font=\scriptsize},colsep=7pt,hlines,vlines}
            \textbf{Scale}&\textbf{GNSS}&\textbf{RTK}&\textbf{\textit{t}-test}\\
            \textbf{SUS} & {M~=~81.7\\SD~=~9.74} & {M~=~74.4\\SD~=~12.0}&{t(51) = 2.45,\\p~=~0.018}\\
            {\textbf{HARUS} (manipulability)} & {M~=~76.7\\SD~=~13} & {M~=~68.1\\SD~=~16.1}&{t(51) = 2.13,\\p~=~0.038}\\
            {\textbf{HARUS} (comprehensibility)} & {M~=~78.3\\SD~=~11.3} & {M~=~74.9\\SD~=~12.9}&{t(51) = 1.01,\\p~=~0.318}\\
            {\textbf{UEQ} (attractiveness)} & {M~=~1.72\\SD~=~0.7} & {M~=~1.1\\SD~=~0.98}&{t(51) = 2.65,\\p~=~0.011}\\
            {\textbf{UEQ} (perspicuity)} & {M~=~2.02\\SD~=~0.64} & {M~=~1.45\\SD~=~0.92}&{t(46.7) = 2.61,\\p~=~0.012}\\
            {\textbf{UEQ} (efficiency)} & {M~=~1.24\\SD~=~0.85} & {M~=~0.85\\SD~=~0.94}&{t(51) = 1.58,\\p~=~0.121}\\
            {\textbf{UEQ} (dependability)} & {M~=~1.17\\SD~=~0.68} & {M~=~1.02\\SD~=~0.62}&{t(51) = 0.87,\\p~=~0.39}\\
            {\textbf{UEQ} (stimulation)} & {M~=~1.84\\SD~=~0.84} & {M~=~1.31\\SD~=~1.11}&{t(51) = 1.93,\\p~=~0.059}\\
            {\textbf{UEQ} (novelty)} & {M~=~1.8\\SD~=~0.85} & {M~=~1.21\\SD~=~0.89}&{t(51) = 2.45,\\p~=~0.018}\\
        \end{tblr}
        \caption{Usability results by group and \textit{t}-tests.}
        \label{tab:usability-table}
    \end{table}
        
        \subsubsection{SUS}\label{sec:sus}
\begin{figure}[ht!]
            \centering
                \includegraphics[width=1.0\columnwidth]{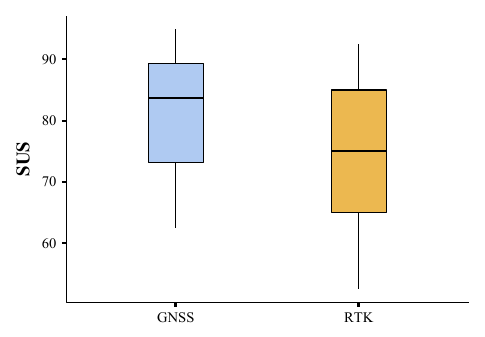}
                \caption{SUS scores by group.}
                \label{fig:sus}
        \end{figure}

        As shown in Figure~\ref{fig:sus}, the mean SUS score for the GNSS group is M~=~81.7 (SD~=~9.74). The mean SUS score for the RTK group is M~=~74.4 (SD~=~12). The results show a significant difference between the two groups (t(51) = 2.45, p~=~0.018). 
        
        \subsubsection{HARUS}\label{sec:harus}
\begin{figure}[ht!]
            \centering
            \includegraphics[width=1.0\columnwidth]{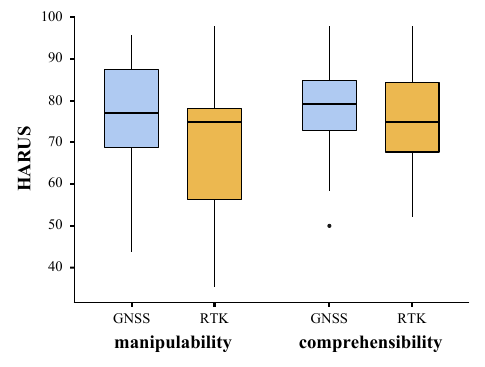}
            \caption{HARUS scores by group.}
            \label{fig:harus-1}
        \end{figure}

        On the \textit{manipulability} scale (indicating ease of handling the AR system), the mean score for the GNSS group is M~=~76.7 (SD~=~13) and that of the RTK group is M~=~68.1 (SD~=~16.1), as shown in Figure~\ref{fig:harus-1}. The results show a significant difference between the two groups (t(51) = 2.13, p~=~0.038). On the \textit{comprehensibility} scale (indicating ease of understanding information presented in the AR interface), the mean score for the GNSS group is M~=~78.3 (SD~=~11.3) whereas the mean score and that of the RTK group is M~=~74.9 (SD~=~12.9). The results \textit{do not} show any significant difference between the two groups (t(51) = 1.01, p~=~0.318). 
        
        \subsubsection{UEQ}\label{sec:ueq}
\begin{figure}[ht!]
            \centering
                \includegraphics[width=1.0\columnwidth]{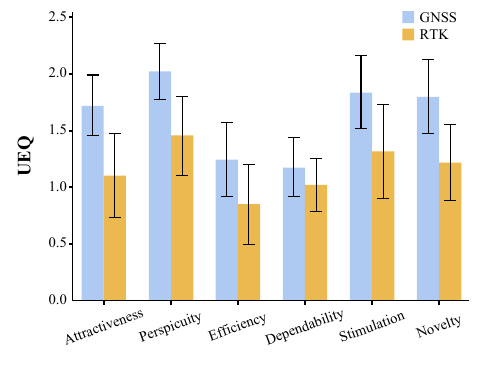}
                \caption{UEQ scores by group.}
                \label{fig:ueq}
        \end{figure}

        As shown in Figure~\ref{fig:ueq}, on the \textit{attractiveness} scale, the mean score for the GNSS group is M~=~1.72 (SD~=~0.7) and that of the RTK group is M~=~1.1 (SD~=~0.98). The results show a significant difference (t(51) = 2.65, p~=~0.011). On the \textit{perspicuity} scale, the mean score for the GNSS group is 2.02 (SD~=~0.64) and that of the RTK group is 1.45 (SD~=~0.92). A Welch \textit{t}-test was used. The results show a significant difference between the two groups (t(46.7) = 2.61, p~=~0.012). On the \textit{efficiency} scale, the mean score for the GNSS group is 1.24 (SD~=~0.85) and that of the RTK group is 0.85 (SD~=~0.94). The results \textit{do not} show any significant difference (t(51) = 1.58, p~=~0.121). On the \textit{dependability} scale, the mean score for the GNSS group is 1.17 (SD~=~0.68) and that of the RTK group is 1.02 (SD~=~0.62). The results \textit{do not} show any significant difference (t(51) = 0.87, p~=~0.39). On the \textit{stimulation} scale, the mean score for the GNSS group is 1.84 (SD~=~0.84) and that of the RTK group is 1.31 (SD~=~1.11). The results \textit{do not} show any significant difference (t(51) = 1.93, p~=~0.059). On the \textit{novelty} scale, the mean score for the GNSS group is 1.8 (SD~=~0.85) and that of the RTK group is 1.21 (SD~=~0.89). The results show a significant difference (t(51) = 2.45, p~=~0.018). 
        
    \subsection{Exploration}\label{sec:exploration-results}
    
        \subsubsection{Distance walked}\label{sec:results-distance-walked} 
\begin{figure}[ht!]
            \centering
            \includegraphics[width=1.0\columnwidth]{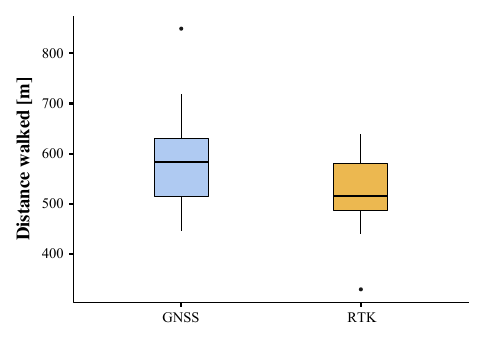}
            \caption{Distance walked by group.}
            \label{fig:distance-walked}
        \end{figure}

As shown in Figure~\ref{fig:distance-walked}, the GNSS group walked an average distance of M~=~586.15 (SD~=~96.24) meters, whereas the RTK group walked an average distance of M~=~525.94 (SD~=~71.9) meters. The results show a significant difference (t(51) = 2.59, p~=~0.013). 
    
        \subsubsection{POIs visited}\label{sec:results-poi-visited} The GNSS group visited an average of M~=~21.09 (SD~=~4.02) POIs, whereas the RTK group visited an average of M~=~19.29 (SD~=~5.87). The results \textit{do not} show any significant difference (t(51) = 1.30, p~=~0.199). 
    
        \subsubsection{Map opened}\label{sec:results-map-opened} The GNSS group opened the 2D map M~=~2.83 (SD~=~2.24) times in average, whereas the RTK group opened it M~=~1.91 (SD~=~2.41) times. The results \textit{do not} show any significant difference (t(51) = 1.44, p~=~0.157). 
        
    \subsection{Focus}\label{sec:focus}
    The GNSS group spend an average M~=~73.3\% (SD~=~9.81) of the time looking at the tablet screen. The RTK group spend an average M~=~69.2\% (SD~=~12.4) of the time looking at the tablet screen. The results \textit{do not} show any significant difference (t(51) = 1.16, p~=~0.251).

\section{Conclusions}\label{sec:conclusions}

The purpose of the study was to assess the impact of geolocation data on the usability of our location-based AR system. To test our hypotheses, we exposed the participants to different geolocation data sources with significantly different accuracies. While we expected RTK data to be more accurate and that it would enable us to improve usability, analysis highlights that it was significantly less accurate and less continuous than GNSS data. This appears to be due to the fact that the embedded GNSS sensor contains filters that preprocess data and remove most of the outliers. In contrast, RTK data purposefully remains “raw”, which is valuable for an advanced user. RTK data accuracy is very efficient when used on an isolated basis (ie. at a 2D map scale), but not particularly suitable for a real-time continuous usage (where location is measured several times per second) on a 1:1, tridimensional scale, at least without any filters applied onto it. Despite this contingency, both the quality and continuity of the geolocation data accuracy the two groups were exposed to was significantly different, which is the essential premise for testing our hypothesis and addressing our research questions. Regarding our main research question, results reveal that the GNSS group, who used the AR application in combination with more accurate and continuous data, reported higher scores in all usability scales, of which five out of nine were statistically significant. This supports our initial hypothesis that poor data accuracy negatively impacts the usability of a location-based AR system. Futures studies should however investigate whether RTK data with proper outlier processing may actually better usability. Our results further highlight that the GNSS group walked more than the RTK group, revealing that the accuracy of geolocation data was partially related to exploration, at least for the quantity indicator. However, due to the manual removal of the outliers–which were significantly more frequent in the RTK group–from the trajectories, the data could be biased. It would be necessary to record a trajectory with both modalities, remove the outliers and observe if there are not significant difference between the measurements to ensure that there are no bias. The comparison on the exploration diversity indicator (amount of POIs visited) was not significantly different. Additionally, although the difference was not significant, the GNSS group opened the 2D map more often than the RTK group in average, suggesting the RTK group could have had more ease exploring. Our results further highlight that there were no significant difference between the ratio of time participants from each group spent interacting with the tablet screen, which would indicate that there is no particular relation between the accuracy of geolocation data and focus. 

Although the two experiments cannot be properly compared, because the tests took place 5 years apart under different conditions, we note that participants spent 69.2\%--73.3\% of the time looking at the tablet screen, which seems to be a meaningful longitudinal progress from the measurement that was made on our 2017 proof-of-concept, where participants interacted with the screen for 88.5~\% of the time~\cite{ingensand_augmented_2018}. While we are not aware of a method to determine the ideal proportion, this measure overall remains an interesting indicator of the importance of the tablet in this type of activity. In a wide review of mobile learning projects, technology was found to dominate the experience in a problematic way in 70\% (28/38) of the cases~\cite{goth_focus_2006}. While using RTK data did not allow us to positively impact the usability of our system, our study however demonstrated the impact of varying geolocation data accuracy on usability and exploration. The immediate benefit of performing this comparative study is for us to define the most suitable conditions of use before offering our system to a young audience, as well as to ensure an adequate overall level of usability. The overall score reported by the GNSS group allows us to qualify the application’s usability as “excellent” according to the SUS adjective scale~\cite{bangor_determining_2009}.

\section{Acknowledgements}\label{sec:acknowledgements}
The authors thank Yoann Douillet for his help with the organization of the tests and the eye tracking data collection. Study participation was voluntary, and written informed consent to publish this paper was obtained from all participants involved in the study. Participants were informed that they could withdraw from the study at any point. The data presented in this study is openly available on Zenodo at \url{https://zenodo.org/record/7845707}. This research was funded by the Swiss National Science Foundation (SNSF) as part of the NRP 77 “Digital Transformation” (project number 407740\_187313) and by the University of Applied Sciences and Arts Western Switzerland (HES-SO): Programme stratégique “Transition numérique et enjeux sociétaux”. The authors declare no conflict of interest. The funders had no role in the design of the study; in the collection, analyses, or interpretation of data; in the writing of the manuscript, or in the decision to publish the results. 

{
	\begin{spacing}{1.17}
		\normalsize
  \bibliography{FOSS4G-references}
	\end{spacing}
}

\end{document}